\documentclass[reprint,superscriptaddress]{revtex4-1}

\usepackage[T1]{fontenc}
\usepackage[utf8]{inputenc}
\usepackage{url}
\usepackage{amsmath}
\usepackage{mathtools}
\usepackage{bbold}
\usepackage{mathrsfs}
\DeclarePairedDelimiter{\abs}{\lvert}{\rvert}
\usepackage{graphicx}
\usepackage{epstopdf} 
\usepackage{wrapfig}
\usepackage{braket}
\usepackage{xcolor}

\begin{document}

\author{Giovanni Pecci}
\affiliation{Univ. Grenoble Alpes, CNRS, LPMMC, 38000 Grenoble, France}

\author{Piero Naldesi}
\affiliation{Univ. Grenoble Alpes, CNRS, LPMMC, 38000 Grenoble, France}

\author{Anna Minguzzi}
\affiliation{Univ. Grenoble Alpes, CNRS, LPMMC, 38000 Grenoble, France}

\author{Luigi Amico}

\affiliation{Quantum Research Centre, Technology Innovation Institute, Abu Dhabi, UAE}

\affiliation{Centre for Quantum Technologies, National University of Singapore, 3 Science Drive 2, Singapore 117543, Singapore}
\affiliation{LANEF Chaire d’excellence, Univ. Grenoble-Alpes $\&$ CNRS, F-38000 Grenoble, France}

\title{Single-particle versus many-body phase coherence in an interacting Fermi gas}

\begin{abstract} 
In quantum mechanics, each particle is described by a complex valued wave-function characterized by amplitude and phase. When many particles interact each other, cooperative phenomena 
give rise to a quantum many-body state with a specific quantum coherence. What is the interplay between single-particle's phase coherence and many-body quantum coherence? 
Over the years, such question has been object of profound analysis in quantum physics.
Here, we demonstrate how the time-dependent interference formed by releasing an interacting degenerate Fermi gas from a specific matter-wave circuit in an effective magnetic field can tell apart the two notions. Single-particle phase coherence, indicated by the first-order correlator, and many-body quantum coherence, indicated by the density-density correlator, are displayed as distinct features of the interferogram. Single particle phase coherence produces spiral interference of the Fermi orbitals at intermediate times. Many-body quantum coherence emerges as long times interference. The interplay between single-particle coherence and many-body coherence is reflected in a stepwise dependence of the interference pattern on the effective magnetic field.
\end{abstract}

\maketitle

Phase coherence is the ability of the quantum wave-function to retain its phase information. 
For free particles, such notion is 
operatively inferred, for example, through the interference pattern in two slits experiments. 
When it comes to be referred to interacting quantum many particles systems, though, the notion of phase coherence is more involved.
While all quantum particles can generally cooperate to establish a macroscopic coherence \cite{blundell2003magnetism,RevModPhys.34.694}, bosons and fermions  display 
two different behaviours.
Bosonic systems can form a Bose-Einstein condensate and be described by a
macroscopic wave-function $\Psi \sim e^{i\phi}$, with a phase $\phi$ coinciding with the single-particle phase. With the spectacular advances on atom trapping and cooling such phase coherence has been studied with unprecedented degree of control and precision of physical conditions. Although cold atoms systems are made of large but finite number of particles ($\sim10^5$),  Bose-Einstein Condensate (BEC) provide a meaningful interference pattern as a result of self-averaging \cite{andrews1997observation,castin1997relative}. {Fundamental limits on fringes visibility are provided by phase diffusion \cite{PhysRevLett.77.3489,PhysRevA.88.063623}}.
Bose Josephson effect is a direct test for BEC coherence \cite{PhysRevLett.95.010402,levy2007ac,cataliotti2001josephson,smerzi1997quantum}. The phase portrait of ultracold bosonic systems has been analysed recently through a series of interferometric experiments (heterodyne phase detection protocol) \cite{eckel2014interferometric,corman2014quench,mathew2015self}, within the emerging field of atomtronics \cite{amico2021roadmap,amico2021atomtronic}. 

For fermionic systems, the Pauli principle
prevents the occupancy of a single quantum level by particles with the same spin, and therefore 
 the many-body coherence is achieved from the coherence of each single particle 
through more complicated mechanisms 
\cite{leggett2006quantum}.
The persistent current \cite{imry2002introduction,bleszynski2009persistent,tinkham2004introduction},  a frictionless flow  occurring e.g.
in metallic or superconducting small rings, provides a characteristic trait of quantum 
coherence in fermionic systems.
  The nature of the many-body phase coherence in degenerate fermions, though,  
  depends on particles interaction.  Ultracold atoms experiments provide an ideal platform to explore these effects, with interactions that can be adjusted from repulsive to attractive cases. \cite{inguscio2007ultra}.
For repulsive interactions, the effects 
of the ferromagnetic correlations have been  observed \cite{massignan2014polarons,PhysRevLett.121.253602}.
For attractive interactions, bound states of Fermi pairs can condense, experiencing the BCS-BEC crossover \cite{strinati2018bcs}. Relevant information on the system coherence in the crossover can be extracted through the study of the Josephson effect \cite{valtolina2015josephson,kwon2020strongly,luick2020ideal}.

\begin{figure}[h!!!]
\includegraphics[width=1\columnwidth]{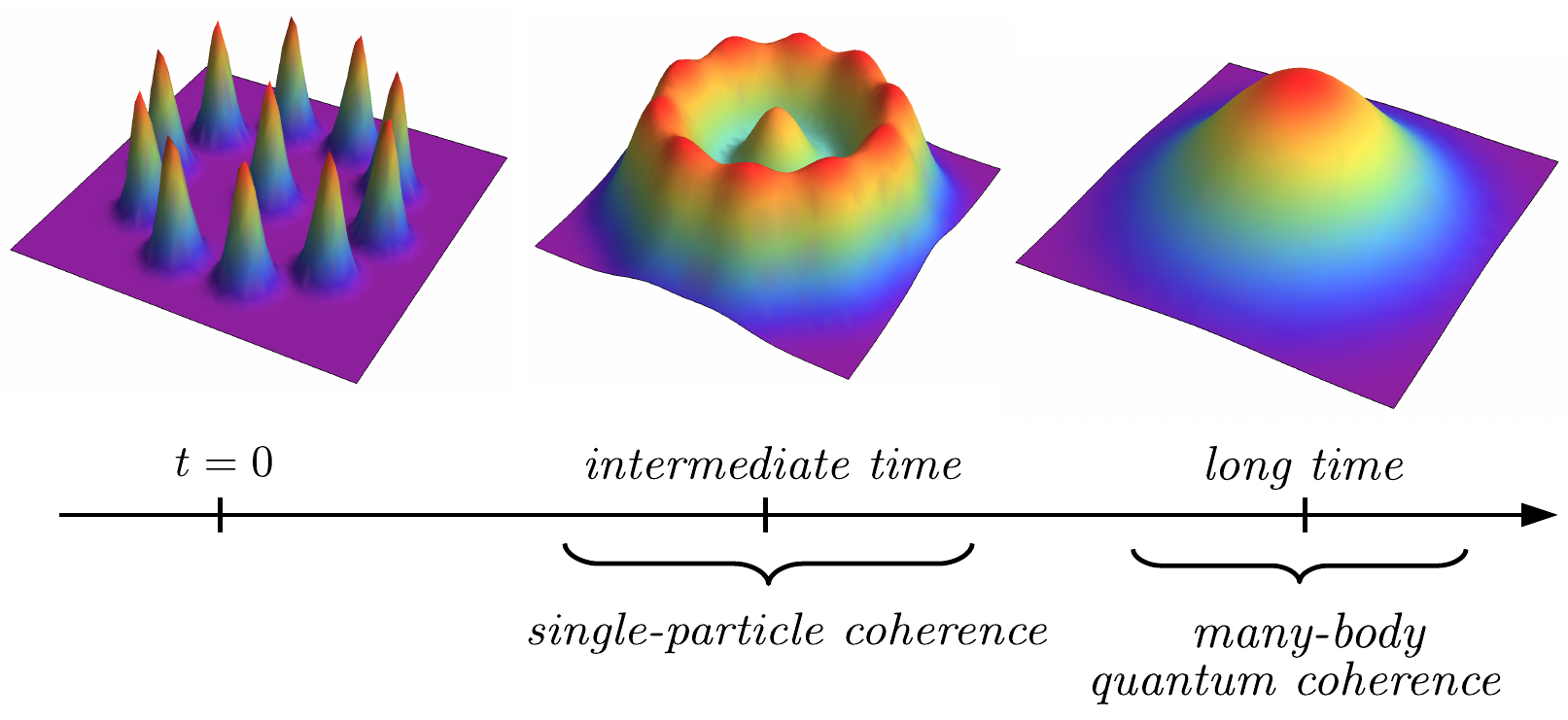}
\caption{Expansion protocol at different times. After the release of the trap, the particles in the center and on the ring interfere. We shall see the single-particle coherence to emerge at intermediate times, while in the long-time limit this protocol provides information about the many-body quantum coherence.}
\label{Protocol}
\end{figure}

\begin{figure}
\includegraphics[width=1\columnwidth]{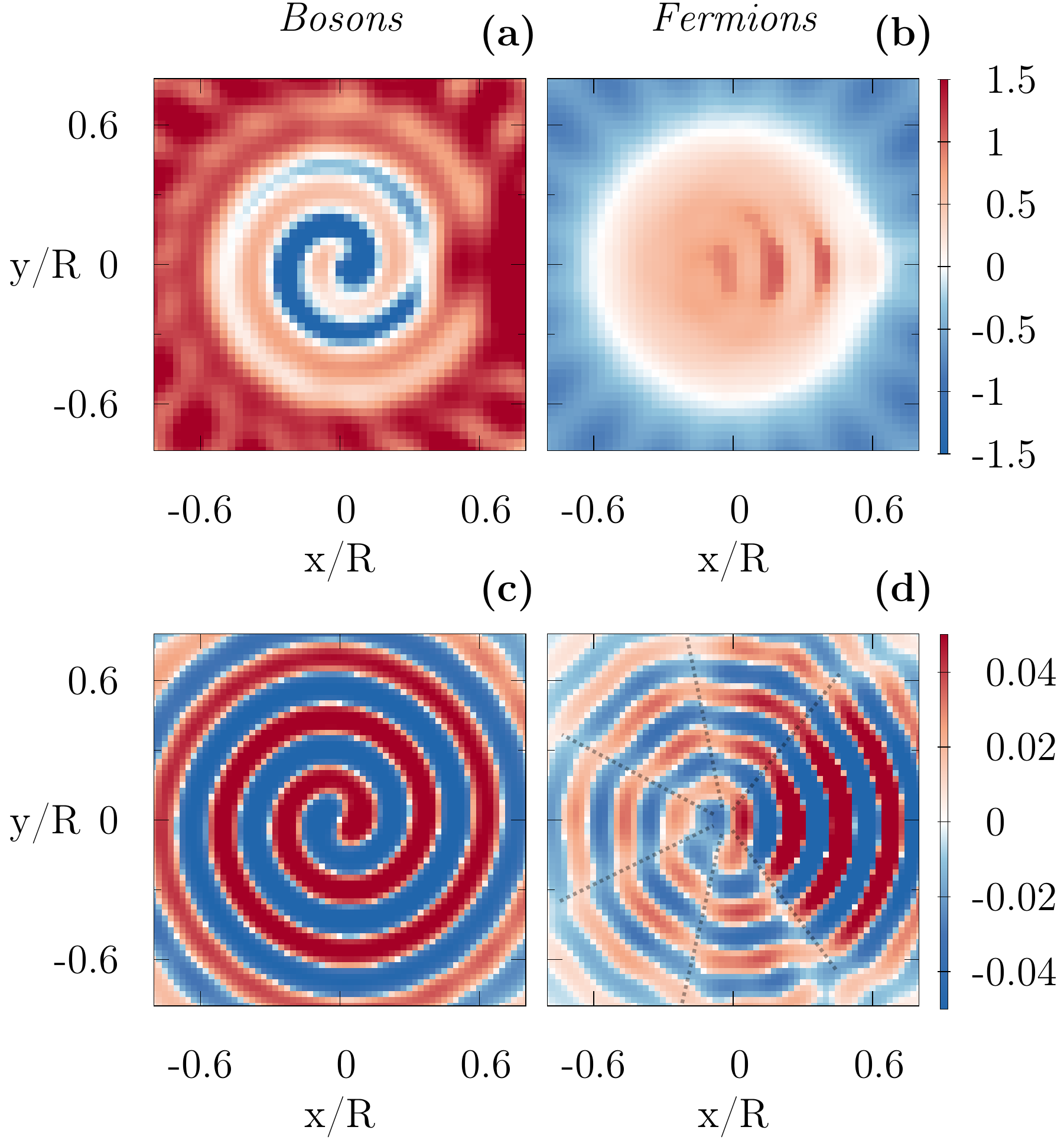}
\caption{Intermediate times density-density correlations. The Bose (left column) and the Fermi (right column) case are compared. In the upper panels, the overall correlation $G(\mathbf{r},\mathbf{r'},t)$ for $\omega_0 t=3$ is presented.
In $(c)$ and $(d)$, the interference $\tilde{G}^{(\text{C},\text{R})}(\mathbf{r},\mathbf{r'},t)$ between the ring and the central site is displayed (see also text). Calculation performed by DMRG on a system of $N=14$ particles on $N_s = 20$ sites, with interaction $U/J=0.2$ and $\tilde{\Omega} = 1.4$. In every plot, $\mathbf{r'} = (R,0)$.}
\label{Full_Interference}
\end{figure}

While the features mentioned above do provide specific aspects for single particle and many-body quantum coherence, their interplay  in interacting many-body systems remains unclear \cite{leggett1991concept,leggett2006quantum}.
In this paper, we operatively track the aforementioned interplay 
through a single protocol, the expansion dynamics of an interacting Fermi gas, that is well within the current state of the art of the cold atoms research field\cite{cai2021persistent,del2022imprinting}. To this end, we study a degenerate interacting Fermi gas confined in a ring-shaped potential and pierced by an effective magnetic flux. Because of the effective magnetic flux, a current is imprinted on the degenerate gas. In analogy with bosonic protocols \cite{haug2018readout,castin1997relative,eckel2014interferometric,corman2014quench,mathew2015self}, such system is let to interfere with a second degenerate gas placed at the center of the ring. In the following, we consider the gas in the center to be composed by two fermions with opposite spins. We stress that the density of the gas in the center does not affect qualitatively our results. Next, we study the entire time evolution of such expansion. 
Despite the similarity in the schemes, we shall see that the fermionic interferograms are markedly different from the bosonic ones. We show that the particles phase coherence emerges in the intermediate times interference images, displaying characteristic dislocations (see dashed lines in Fig.\ref{Full_Interference}d) due to Fermi sphere effects. On the other hand, the many-body coherence, exemplified by pairing correlations and off-diagonal-long-range order (ODLRO), ie long-ranged spatial coherence \cite{RevModPhys.34.694}, emerge at long times in the density-density correlators. The response of the many-body coherence to magnetic field arises as a step-wise dependence of the density-density correlators on the magnetic field. 

\paragraph*{{Model.}}
We consider a gas of $N$ degenerate fermions confined in a ring lattice  of radius $R$ comprised of $N_s$ sites and 
pierced by an effective magnetic flux $\Omega$ induced by an
artificial gauge field. Such effective magnetic flux can be applied in several ways, for example by stirring the gas, by phase imprinting or by two-photon Raman transitions \cite{dalibard2011colloquium}.
The system is described by the Hubbard Hamiltonian in a one-dimensional ($1d$) ring-shaped spatial geometry
\begin{equation}
\small{\hat{\mathcal{H}}_{FH} = -J \sum_{j=1}^{N_s}\sum_{\alpha =\uparrow,\downarrow} \Bigl(e^{i \tilde{\Omega}} \hat c^\dagger_{j, \sigma} \hat c_{j+1, \sigma} + \textit{H.c.} \Bigr) + U \sum_{j=1}^{N_s} \hat n_{j,\uparrow} \hat n_{j,\downarrow} }
\label{Hubbard}
\end{equation}
where $U$ is the particles interaction and $J$ is the tunnel amplitude. The applied gauge field is taken into account through the Peierls phase factors $e^{i\tilde{\Omega}}$, in which $\tilde{\Omega}\doteq \frac{2\pi }{N_s}\frac{\Omega}{\Omega_0} = \frac{2\pi }{N_s}\frac{\phi}{\phi_0}$, with $\Omega_0 = \frac{\hbar}{ m R^2}$ being the typical frequency of the ring and $m$ the particle mass. In analogy with electronic systems, we also introduced $\phi = \pi R^2 \Omega$ as the artificial gauge flux and $\phi_0 = \frac{h}{2m}$ as the flux quantum. We remark that fermions in continuous rings with delta-interaction can be described with Hubbard rings lattices in the small fillings $N/N_s$ regimes \cite{SupplMaterial}.

For $U>0 $, the ground-state many-body wave-function is made of extended states characterized by real wave-momenta \cite{andrei1995integrable}. 
For $U<0$, the ground state is characterized by bound pairs, with complex wave-momenta. For weak attractions, the ground state of the system is a BCS-like state with the wave-function of the pair decaying on distance larger then the mean interparticle separation \cite{fuchs2004exactly,marsiglio1997evaluation,guan2013fermi}.
For stronger attractions, the bound states are formed by tightly bound particle pairs. 
The persistent currents in Hubbard models was studied for repulsive interactions  in \cite{PhysRevLett.95.063201}  and for attractive interactions 
in \cite{Pecci2020probing}. 

In this work, we study the co-expansion of the initially ring-trapped Fermi gas and the one located at the ring's center, ie two fermionic atoms with opposite spin, after the two confining potentials are suddenly switched off simultaneously \cite{Roscilde_2016,haug2018readout}. We integrate the density profiles along the $z$ axis and focus on the dependence on $\mathbf{r}=(x,y)$ in the ring plane. The field operator of the whole system (ring plus centre) is
\begin{equation}
\hat \Psi_\alpha ({\mathbf r},t)=w_\text{C}({\mathbf r},t) \hat c_{\text{C},\alpha}+\sum_{j=1}^{N_s} w_j({\mathbf r},t) \hat c_{j,\alpha}, 
\end{equation}
where $\alpha = \uparrow, \downarrow$, $\text{C}$ indicates the central site and
\begin{equation}
w_\Lambda(\mathbf{r},t) = \frac{\exp\{-(\mathbf{r} -\mathbf{r}_\Lambda)^2/(2\sigma^2(1+ i \omega_0 t))\}}{\sqrt{\pi}\sigma(1 + i \omega_0 t)} \quad \Lambda=\{ j,\text{C}\}
\end{equation}
is the time-dependent Wannier function centered at position $\mathbf{r}_\Lambda$ in the Gaussian approximation \cite{wannier1978transforming}, $\omega_0$ being the frequency of each lattice well in the harmonic approximation. This allows for an explicit solution for the dynamics of the field operator following a sudden turn-off of the lattice \cite{PhysRevLett.94.240404,chiofalo2000collective}.
In the experiment with weakly interacting bosonic condensates, a spiral interferogram emerges in a single co-expansion.
However, in our theoretical approach, the particle density 
$\braket{ \hat n({\mathbf r},t)}\doteq \sum_{\alpha = \uparrow, \downarrow} \braket{\hat n_\alpha({\mathbf r},t)} \doteq \langle \hat \Psi^\dagger_\alpha({\mathbf r},t) \hat \Psi_\alpha({\mathbf r},t) \rangle$ is reconstructed as an expectation value, corresponding to an average over different realizations of the co-expansion protocol.
Since each co-expansion is characterized by a well defined, yet randomly distributed relative phase between 
the particles released from the ring and from the central site, 
the interference 
pattern is washed out in $n(\mathbf{r},t)$. We shall see that non trivial phase information on the system is captured by the density-density correlator: 
\begin{align}
G({\mathbf r},{\mathbf r'};t) &\doteq \sum_{\alpha,\beta = \uparrow, \downarrow}G_{\alpha, \beta}({\mathbf r},{\mathbf r'};t) \ \ \ \text{where}\\ 
G_{\alpha, \beta}({\mathbf r},{\mathbf r'};t) & \doteq \langle \hat \Psi_\alpha^\dagger({\mathbf r},t) \hat \Psi_\alpha({\mathbf r},t) \hat \Psi_\beta^\dagger({\mathbf r'},t)\hat \Psi_\beta({\mathbf r'},t)\rangle.
\end{align}
For our combined centre-ring system, such correlation can be broken down as 
$ G({\mathbf r},{\mathbf r'},t)=G^{(\text{C},\text{C})}+G^{(\text{R},\text{R})}+G^{(\text{C},\text{R})}$, where $\text{C}$ refers to central site and $\text{R}$ to the ring.
Since the ring lattice and the central site are disconnected, we have 
$|\psi(t\!=\!0)\rangle \!=\! |\psi(t\!=\!0)\rangle_{\text{C}} \otimes |\psi(t\!=\!0)\rangle_{\text{R}}$. Assuming a free expansion for $t\ge 0$, the ring-centre correlations read $G^{(\text{C},\text{R})} ({\mathbf r},{\mathbf r'},t)=\braket{\hat n^{(\text{R})}({\mathbf r},t)}\braket{ \hat n^{(\text{C})}({\mathbf r'},t)}+\braket{\hat n^{(\text{R})}({\mathbf r'},t) }\braket{\hat n^{(\text{C})}({\mathbf r},t)}+\tilde{G}^{(\text{C},\text{R})}({\mathbf r},{\mathbf r'},t)$, where
\begin{equation}
\tilde{G}^{(\text{C},\text{R})}({\mathbf r},{\mathbf r'},t)=\sum_{\alpha} \sum_{i,j} I_{i,j} ({\mathbf r},{\mathbf r'},t) \left( \delta_{ij} -\langle \hat c^\dagger_{i,\alpha} \hat c_{j,\alpha} \rangle \right)
\label{interference}
\end{equation}
with $I_{ij}({\mathbf r},{\mathbf r'},t) \doteq w^*_\text{C}({\mathbf r},t) w_\text{C}({\mathbf r'},t) w_i^*({\mathbf r'},t) w_j({\mathbf r},t) $.
From Eq.~(\ref{interference}) we see that the interference between the center and the ring only affects the particles belonging to the same spin species and depends on the first-order correlator $\langle \hat c^\dagger_{i,\alpha} \hat c_{j,\alpha} \rangle$. 
We note that the $\tilde{G}^{(\text{C},\text{R})}$ correlator can be accessed by measuring the full correlator $G({\mathbf r},{\mathbf r'},t)$, and by subtracting the ring-ring  $G^{(\text{R},\text{R})}$ and center-center contributions $G^{(\text{C},\text{C})}$,  as well as the  densities $\braket{ \hat n^{(\text{R})}({\mathbf r},t)}$,$\braket{ \hat  n^{(\text{C})}({\mathbf r},t)}$, measured independently by eliminating either the ring lattice or the central site.

In our approach, we monitor the complete correlator $G({\mathbf r},{\mathbf r'},t)$, the interference term $\tilde{G}^{(\text{C},\text{R})}$ Eq.~(\ref{interference}) and the spin resolved correlator $G_{\uparrow,\downarrow}$\cite{PhysRevA.97.063613}. 
We shall see that at intermediate times $\tilde{G}^{(\text{C},\text{R})}$ contains direct information on the single-particle phase coherence. Instead, at long times, $G_{\uparrow,\downarrow}({\mathbf r},{\mathbf r'},t)$ probes the many-body phase coherence in momentum space with $\mathbf{k}= m \mathbf{r}/\hbar t_{exp}$, $\mathbf{k'}= m \mathbf{r'}/\hbar t_{exp}$ with 
$t_{exp}$ being the expansion time. 
In the present work we employ  DMRG simulations to obtain
the correlator matrix. For a lattice with $N_s$ sites, this accounts to calculate $N_s^4$ terms.

\begin{figure}
\includegraphics[width=1\columnwidth]{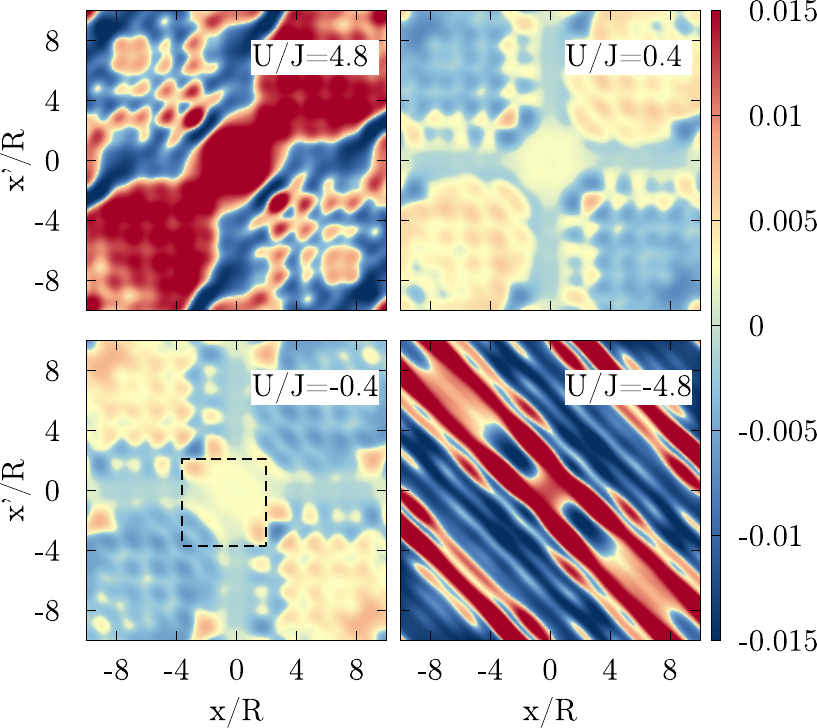}
\caption{Long times connected density-density correlator. The correlation for ${\mathbf r}=(x,0)$, ${\mathbf r'}=(x',0)$ and $t_{exp}= 100 ~\omega_0^{-1}$ is evaluated.
For weak attractive interaction we observe BCS-like correlations at $x=-x'\simeq \hbar t_{exp}k_{F}/m $. At strong attractive interactions tightly bound pairs are revealed by the enhancement of the correlations along the whole $x = -x'$ diagonal. In the bottom left panel, the square indicates the size of the Fermi sphere at $x,x'=\pm \hbar t_{exp}k_{F}$. The calculations are performed using the DMRG method with $N=14$, $N_s=20$ and $\tilde{\Omega}=1.4$. }
\label{Partial_Interference}
\end{figure}

\paragraph*{{Bosons versus fermions.}} For 
bosons, the spiral interference pattern arises because of the simple coupling between the effective gauge field and the phase of the Bose condensate. The quantized circulation reflects the effective magnetic flux quantization.
The complete phase structure of the bosonic field emerges as a characteristic spiral interferogram in the expanding density \cite{eckel2014interferometric,corman2014quench,mathew2015self} as well as in the interference term $\tilde G(\mathbf r,\mathbf{r'};t)$ \cite{haug2018readout} (see Fig.~2 c). 

For fermionic systems, the relation between the imparted phase and the density-density correlator is more involved. The difference traces back to the symmetry properties of the many-body wave-functions of the two systems resulting in a different momentum distribution. Bosonic wave-functions yield a momentum distribution peaked at $k=0$ \cite{moulder2012quantized,wright2013threshold,PhysRevLett.110.025302}, while Fermi systems are characterized by a a broader momentum distribution.
Then, when fermionic particles are put in motion by an effective magnetic flux, each momentum component of the distribution is characterized by a different phase factor. As a consequence, phases recombination occurs, and the time-of-flight image of the density results to be suppressed at $|\mathbf{k}|=0$ only after half of the Fermi sphere is displaced by the effective magnetic flux \cite{SupplMaterial}. 
In $G({\mathbf r},{\mathbf r'};t)$ , at intermediate expansion times, specific dislocations are found in the interference pattern (see Fig.2 d), deforming the smooth spiral-like picture we observe in the bosonic case (Fig.2 c). This is again due to the distinct particle orbitals characterizing the fermionic state. For $U=0$, such orbitals are strictly single-particle, and each of them yields a spiral-like interference \cite{SupplMaterial}. The dislocations, just $N_{\uparrow}-1$ (or equivalently $N_\downarrow-1)$ in number, are due to the interference of the $N_{\uparrow} (N_{\downarrow})$ independent orbitals. Remarkably, the dislocations are clearly visible at small and moderate interactions. By increasing interactions we find that dislocations disappear \cite{SupplMaterial}, revealing that the system cannot be described in terms of independent quasi-particles.

\begin{figure}
\includegraphics[width=1\columnwidth]{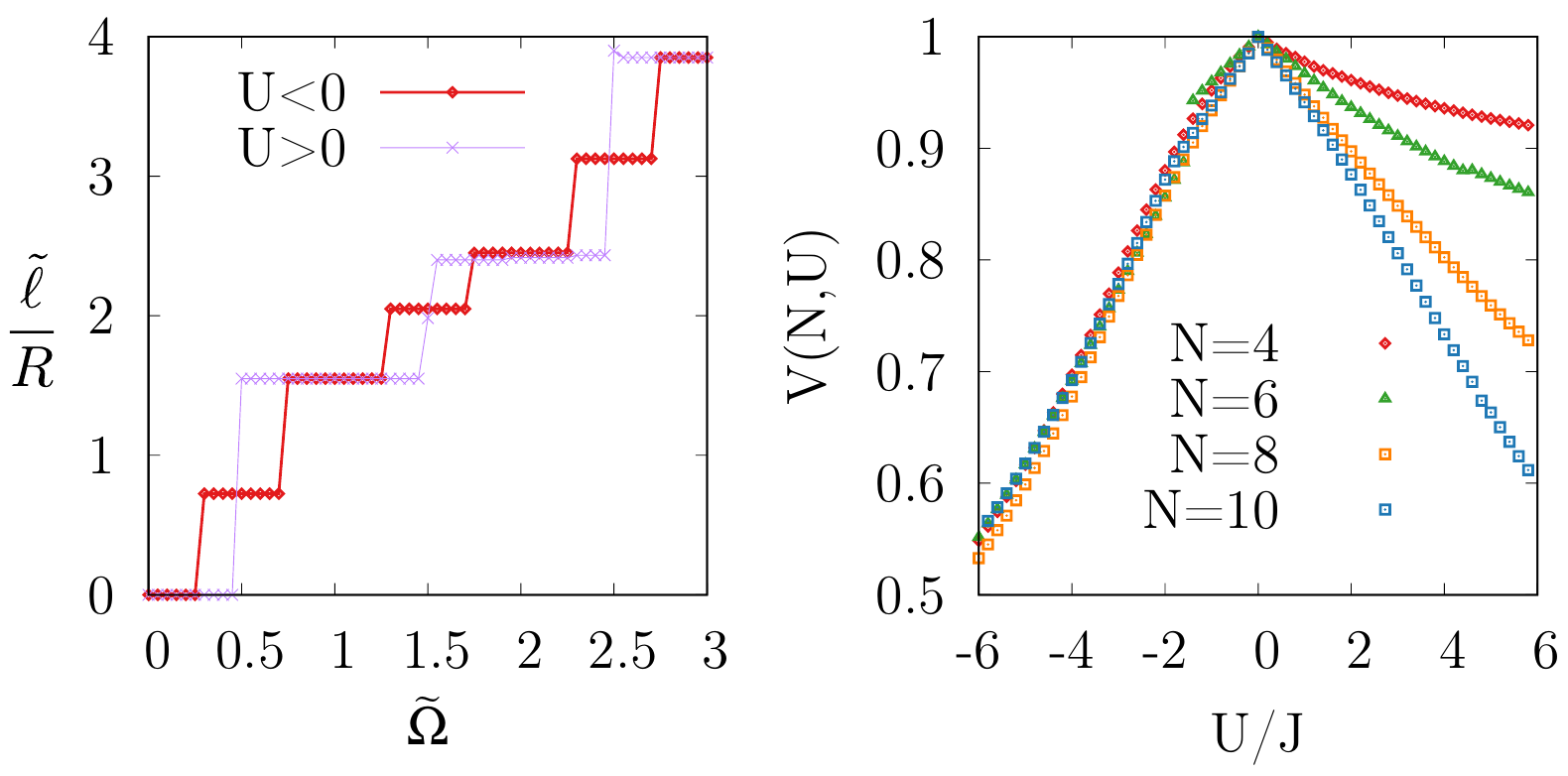}
\caption{Left panel: displacement $\tilde{\ell}/R$ of the correlation peak of Fig.\ref{Partial_Interference} as a function of the flux for $N=6$, $U/J=-5.4$ and $U/J=0.5$. Right panel: visibility of the correlation peak $V(N,U)$, as a function of interactions for $\tilde{\Omega}=0$ and various values of number of particles. 
We observe a markedly different $N$ scaling for repulsive or attractive interactions. In both panels $N_s=10$.}
\label{Fig_steps}
\end{figure}

\paragraph*{{Repulsive vs. attractive interactions.}}
In the following, we demonstrate that the long time expansion of our setup allows us to access to many-body coherence through the connected spin-resolved 
correlator $G_{\uparrow,\downarrow}({\mathbf r},{\mathbf r'};t_{exp}) - \braket{\hat n_{\uparrow}({\mathbf r},t_{exp})}\braket{\hat n_{\downarrow}({\mathbf r'},t_{exp})}$. Notably, we consider the connected part to eliminate the background from the total correlations.
We shall see that the different nature of the many-body state for repulsive and attractive interactions is clearly reflected in our pictures (see Fig. \ref{Partial_Interference}). We note that, within our model, the presence of a reference site does not affect the correlator $G_{\uparrow\downarrow}$, since the coupling between the ring and the center is nonzero only for particles with the same spin.

For $U>0 $ the ground state of the system is made of itinerant correlated particles. Therefore, the interferograms reflect the phase pattern imparted by the effective magnetic flux putting the system in a coherent motion. 
By the analysis of the long-time behavior of $G_{\uparrow\downarrow}({\mathbf r},{\mathbf r'};t)$, at increasing $U$ we find that, for fixed $y$ and $y'$, the correlation has a clear symmetry with respect to $x=x'$ (corresponding in momentum space to $k=k'$), reflecting the tendency of the system to approach magnetic order at large $U/J$ \cite{tasaki2020physics,PhysRevA.79.013609}. The correlations are found to be displaced by a discrete amount as a function of the applied effective magnetic flux, reflecting the quantized particle circulation. This can be observed by looking at the displacement $\tilde{\ell}/R$ of the origin $x=x'=0$ in the direction $x=x'$ as a function of the applied flux (see Fig.~\ref{Fig_steps}). 

For $U<0$, the system is characterized by the off-diagonal (quasi) long-range order 
due to fermionic 
pairing \cite{RevModPhys.34.694}. 
In contrast with repulsive cases, for $U<0$ the correlator $G_{\uparrow\downarrow}({\mathbf r},{\mathbf r'};t)$ displays a marked structure along the whole anti-diagonal $x=-x'$, reflecting the formation of pairs of smaller and smaller size at increasing $|U|$. We found that in the BCS regime the pairs correspond to enhanced $({\mathbf k_F},\mathbf{-k}_F)$ correlations at the Fermi sphere i.e. for wave-vector $|\mathbf k|=k_F$. On the other hand, at larger interactions, pairs of small spatial size are formed and correlations along the whole antidiagonal $(\mathbf k,\mathbf{-k})$ are predicted. 
Such approach is in line with \cite{altman2004probing,staudenmayer2008density} and further analysed in the Supplemental Material \cite{SupplMaterial}. 
Despite our system can be of small size, the antidiagonal correlations features clearly emerge in our expansion protocol (at long times), thus allowing to probe the nature of pairing in the whole BCS-BEC crossover.  The formation of pairs is also reflected in the halving of the period of $\tilde \ell$ as a function of the flux. This is due to the doubling of the mass of the components of the system when atoms are bound in diatomic molecules, which reduces by a factor two the value of the flux quantum. Notably, an equivalent phenomenon occurs in superconducting rings \cite{ByersYang}.

Furthermore, we find that the landscape along the anti-diagonal depends on the number of particles with a markedly different scaling for $U>0$ and $U<0$. In line with Yang's criterion for the off-diagonal-long-range order \cite{RevModPhys.34.694}, we find that for attractive interactions both the maxima and the minima of the momentum correlator along the antidiagonal scale the same way with $N$. Indeed, in the presence of quasi-ODLRO the momentum correlator is dominated by the pair-pair correlations, and scales as $N^\alpha$ with $0<\alpha<1$ for any wave-vector $k$ \cite{SupplMaterial}. 
For $U>0$, instead, the maxima of the momentum correlator are independent on particle number, while the minima 
increase with $N$. As a result, the visibility defined as
\begin{equation}
\small{V(N,U) = \frac{\text{Max}[G_{\uparrow,\downarrow}(x,-x;t)] -\text{Min}[G_{\uparrow,\downarrow}(x,-x;t)]}{\text{Max}[G_{\uparrow,\downarrow}(x,-x;t)] +\text{Min}[G_{\uparrow,\downarrow}(x,-x;t]}}
\end{equation}
and presented in Fig.\ref{Fig_steps}, is independent of $N$ for $U<0$ and decreasing with $N$ for $U>0$. We note that the property clearly emerges already at small $N$, providing a further evidence that ring geometries are well suited for minimizing finite size effects \cite{fisher1972scaling}.

\paragraph*{{Conclusions.}}
In this work, we demonstrated how the interplay between the 
single-particle's phase coherence and the many-body 
phase coherence of an interacting Fermi gas can be probed with a single protocol inspired by heterodyne phase detection schemes employed in cold atoms laboratories. We described the ring-trapped gas through the Hubbard model with the local interaction $U$ ranging from positive to negative values. 
We analysed the dynamics of the density-density correlators. 
They can be accessed, for example, in cold atoms experiments through state-of-the-art processing of the particles expansion.
We note that also continuous (no lattice) ring-shaped degenerate gases can be accessed by our theory in the dilute lattice limit \cite{SupplMaterial}. 
For our protocol, {\it we demonstrated that particle's phase information emerges in the intermediate times interference of the expansion; the many-body phase coherence can be tracked at longer times}. 

For Fermi systems the effective magnetic flux imparts the phase winding on a broad momentum distribution. We have shown 
that the relevant information of the phase of fermionic particles can be traced in the response of the orbitals. This analysis can be carried out in our protocol by suitably extrapolating the ring-center correlations from the total correlations (see Fig.\ref{Full_Interference}). As remarkable spin-off of our analysis on the single orbital interference, we note that our results grants the access to the number of particles $N_{\uparrow} (N_{\downarrow}$). Our prescription is that the number of dislocations obtained in our interferogram Fig.\ref{Full_Interference}d is just $N_{\alpha}-1$ with $\alpha=\uparrow$ or $\downarrow$. 

The proposed protocol provides a step forward towards the realization of the full counting statistics for the system's particle density fluctuations \cite{belzig2007density,staudenmayer2008density}.

For repulsive interactions, we found that the resulting long time image displays enhanced correlations along the diagonal $k=k'$ (reflecting the magnetic ordering). By the application of the effective magnetic flux, the position of the peaks results to be displaced in discrete steps, reflecting the quantized circulation of current along the ring.
For attractive interactions a clear broad anti-diagonal $k=-k'$ surfaces in the expansion. Such feature emerges at $k=k_F$ because of the fermionic pairing, leading to a many-body quantum coherence of the BCS type \cite{altman2004probing,staudenmayer2008density,SupplMaterial}. We find that anti-diagonal correlations arise also for strong attraction in which the pairs are tightly bound; in this case the peaks dissolve on a broad interval of $k$. As  counterpart of the effect found for repulsive fermions, the position of the anti-diagonal results to be displaced by the effective magnetic flux in a quantized fashion. The quantitative analysis shows that for repulsive/attractive interactions the visibility of the anti-diagonal correlations is characterized by a markedly different dependence on the number of particles. Such effect reflects the Yang's off-diagonal-long-range order scaling of the two-body density matrix.

With our work, we bring conceptually relevant aspects of many-body physics to the domain of what can be operatively  tested. Our analysis is timely with the current stage of cold atoms quantum technology: persistent current in toroidal cold fermionic atoms has been achieved in \cite{cai2021persistent,del2022imprinting} and mesoscopic pairing was experimentally analysed in \cite{holten2021observation}. 

\paragraph*{Acknowledgements}
We thank M. Albert, T. Haug, T. Leggett and C. Salomon for discussions. 
The Grenoble LANEF framework ANR-10-LABX-51-01 are acknowledged for their support with mutualized infrastructure.

\bibliography{PhaseCoherence.bib}

\clearpage
\onecolumngrid

\section*{Supplemental Material for "Single-particle versus many-body phase coherence in an interacting Fermi gas"}

\section{General definitions and case  of a free fermion gas}
In this section we recall the general definitions used in the main text and provide analytical results for the case of a free fermion gas. 

We are interested in evaluating 
\begin{equation}
G(\mathbf{r},\mathbf{r'},t) =\sum_{i,j,n,m=0}^{N_s} w^*_i(\mathbf{r},t)w_j(\mathbf{r},t)w^*_n(\mathbf{r'},t)w_m(\mathbf{r'},t)   \langle \hat c_i^\dagger \hat c_j \hat c_n^\dagger \hat c_m \rangle ,
\label{correlator_first}
\end{equation}
where $\hat c_i$ and $\hat c_i^\dagger$ are the fermionic annihilation and creation operators and we have indicated by $i=0$ the central site. It should be noted that, since in the following we will consider just one spin species, we dropped the spin indices $\sigma$. The Wannier functions $w_j(\mathbf{r})$ are defined as
\begin{equation}
w_j(\mathbf{r},t) = \frac{1}{\sqrt{\pi}\sigma } \frac{1}{(1 + i \omega_0 t)} \exp\Bigl\{-\frac{(\mathbf{r} -\mathbf{r}_j)^2}{2\sigma^2(1+ i \omega_0 t)}\Bigr\},
\label{wannier}
\end{equation}
where $\sigma=\sqrt{\hbar/m \omega_0}$ and $\mathbf{r}_j$ are the initial width and the center of the $j$-th Wannier function respectively, with $\omega_0$ being the frequency of the bottom of each lattice well in the harmonic approximation. 
As remarked in the main text, imposing that the center and the ring are totally decoupled, the interference between the ring and the central site is fully encoded in
\begin{equation}
\tilde{G}^{(\text{C},\text{R})}(\mathbf{r},\mathbf{r'},t) = \sum_{k,j}^{N_s} w^*_0(\mathbf{r},t)w_j(\mathbf{r},t)w^*_k(\mathbf{r'},t)w_0(\mathbf{r'},t)(\delta_{kj}-\langle \hat c_k^\dagger \hat c_j\rangle),
\label{withdelta}
\end{equation}
where we used the relation $\langle \hat c_0^\dagger \hat c_0 \rangle = 1$.

In order to isolate the term explicitly which yields the spiral interferograms,
in the following we subtract the term depending on the Kronecker delta
\begin{equation}
\tilde G_0(\mathbf{r},\mathbf{r'},t)=\sum_{k,j}^{N_s} w^*_0(\mathbf{r},t)w_j(\mathbf{r},t)w^*_k(\mathbf{r'},t)w_0(\mathbf{r'},t)\delta_{kj}.
\end{equation}
Such expression corresponds to the expectation value of the one-body density matrix of a non-interacting Fermi gas for a completely filled lattice, when the hopping between different sites is inhibited by the Pauli principle and it forms a band insulator. Its value can be obtained from  the knowledge of the lattice properties.  This term can be experimentally obtained by a separate measurement of the correlator at large filling.  The knowledge of this term would hence allow to  highlight at best the spiral interferogram.

For non-interacting fermions, the one-body correlator $\langle \hat c_k^\dagger \hat c_j \rangle$ can be determined by solving the non-interacting Schroedinger equation under  periodic boundary conditions. An explicit calculation yields:
\begin{equation}
\langle \hat c_k^\dagger \hat c_j \rangle = \sum_{\{n\}} e^{-i \frac{2\pi}{N_s}(j-k)n},
\label{correlation_matrix}
\end{equation}
where $\{n\}$ is the set of integer quantum numbers labelling the energy levels of the Fermi sphere, whose value depends on the applied flux $\Omega$.

Hence, the relevant noise correlator for spiral interferograms reads:
\begin{equation}
\tilde{G}(\mathbf{r},\mathbf{r'},t)= G(\mathbf{r},\mathbf{r'},t) -\tilde{G}_0(\mathbf{r},\mathbf{r'},t)- \braket{\hat n^{(\text{R})}(\mathbf{r},t)} \braket{ \hat n^{(\text{C})}(\mathbf{r'},t)}- \braket{ \hat n^{(\text{C})}(\mathbf{r},t)} \braket{ \hat n^{(\text{R})}(\mathbf{r'},t)} - G^{(\text{C},\text{C})} (\mathbf{r},\mathbf{r'},t)- G^{(\text{R},\text{R})}(\mathbf{r},\mathbf{r'},t),
\label{corr1}
\end{equation}
where we introduced:
\begin{align}
&G^{(\text{C},\text{C})} \equiv \lvert w_0(r,t) \rvert^2 \lvert w_0(r',t) \rvert^2 (\langle \hat c_0^\dagger \hat c_0\rangle)^2 , \notag \\
&G^{(\text{R},\text{R})} \equiv \sum_{i,j,n,m=1}^{N_s}w^*_k(\mathbf{r},t)w_j(\mathbf{r},t)w^*_n(\mathbf{r'},t)w_m(\mathbf{r'},t)   \langle \hat c_i^\dagger \hat c_j \hat c_n^\dagger \hat c_m \rangle 
\end{align}
respectively as the correlations functions only of the central site and only among the particles on the ring.

By using Eq.~(\ref{wannier}) in Eq.~(\ref{corr1}) we readily obtain:
\begin{align}
&\tilde G(\mathbf{r},\mathbf{r'},t)= \notag \\
&= - \abs{\textit{A}(t)}^4 \ e^{-\frac{\sigma^2}{2b^2(t)}(r^2+r'^2)}e^{-i\frac{\hbar t}{mb^2(t)}(r^2-r'^2)}\sum_{kj}e^{-\frac{1}{2 \sigma^2 b^2(t)}((r-r_j)^2 +(r'-r_k)^2)}e^{i\frac{\hbar t}{mb^2(t)}((r-r_j)^2-(r'-r_k)^2)} \sum_{\{n\}} e^{-i \frac{2\pi}{N_s}(j-k)n} \notag \\
&\equiv - \sum_{\{n\}} I_n(\mathbf{r})I_n^*(\mathbf{r'})
\label{inter}
\end{align}
where $b(t) = \sqrt{1 + \omega_0 ^2 t^2}$ 
and $\textit{A}(t)$ is the complex time-dependent amplitude of the Wannier functions. In the last identity we have separated the summation over the two indices: once $r'$ is fixed, Eq.~\ref{inter}) is a superposition of single-orbital functions $I_n(r)$, each weighted with a $n$-dependent coefficient. We plot $\tilde{G}^{(\text{C},\text{R})}(\mathbf{r},\mathbf{r'},t) = \tilde{G}(\mathbf{r},\mathbf{r'},t) + \tilde{G}_0(\mathbf{r},\mathbf{r'},t)$ in the bottom right panel of Fig.$2$ of the main text. Furthermore, we display the full correlator $G(\mathbf{r},\mathbf{r'},t)$. 
Finally, we can make a comparison with the case of non-interacting bosons. In this case, a single quantum state, labelled by a quantum number $n$, is macroscopically occupied. As a consequence, the full correlator is proportional to the function $ I_n(\mathbf{r})I_n^*(\mathbf{r'})$, yielding a perfect spiral interference pattern whose number of branches is equal to $n$.

\begin{figure}
\includegraphics[width=0.7\textwidth]{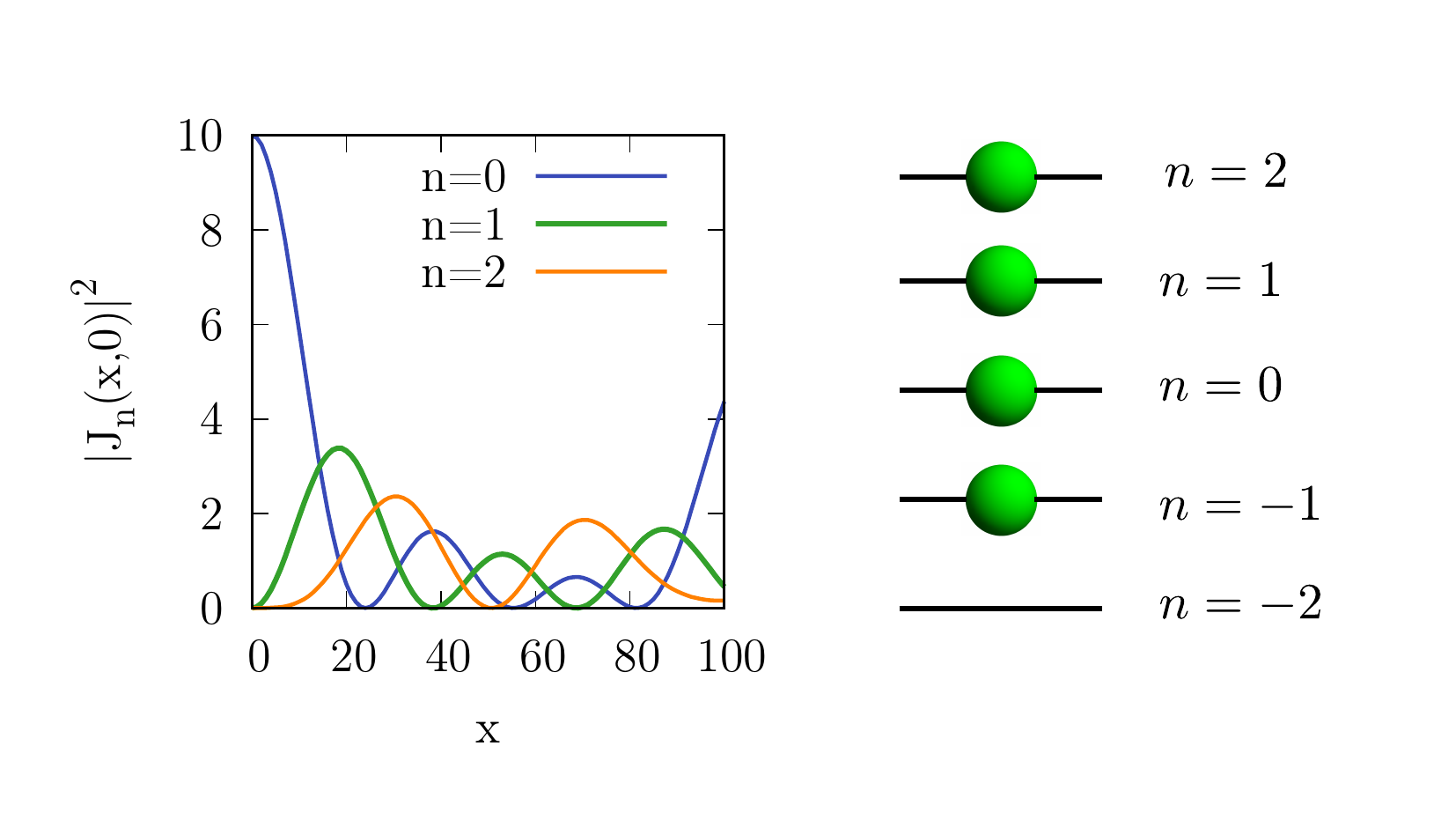}
\centering
\caption{On the left, profile $\big\lvert J_n(x,0)\big\rvert ^2$ for different $n$. We see that when $n=0$ the function has a finite value in the origin $x=y=0$. As $n$ increases, the function collapses to zero in this point. On the right, we show the configuration of the quantum numbers for $N=4$ spinless fermions. In the ground state, they are selected in order to minimize the value of the current.}
\label{J_n}
\end{figure}

\section{Time of flight}
We now focus on the momentum distribution of the particles on the ring, that can be addressed through time of flight imaging using the same protocol proposed in the main text.
Such observable can be accessed by measuring the density distribution of the expanding gas at long times. Indeed 
 by taking the limit $t \to \infty$ in Eq.(\ref{wannier}), one gets $w_j(\mathbf{r},t)\propto e^{i {\bf k(r)} \cdot {\bf r_j}}$ where ${\bf k(r)} = \frac{m}{\hbar t}{\bf r}$. Hence, one has
$\lim_{t \to \infty} \braket{ \hat n({\bf r},t)} \propto  \braket{\hat n({\bf k})} $, where 
the two-dimensional momentum distribution is defined as
\begin{equation}
 \braket{ \hat n({\bf k})} \propto \sum_{i,j}^{N_s} e^{i {\bf k} \cdot ({\bf r_i} - {\bf r_j})} \langle \hat c_i^\dagger \hat c_j \rangle
\end{equation}
where ${\bf r}_i$ and ${\bf r}_j$ are the position of the lattice sites $i$ and $j$. Using the one-body correlation function evaluated in Eq. (\ref{correlation_matrix}), one gets:
\begin{equation}
 \braket{\hat n({\bf k})} \propto \frac{1}{N_s}\sum_{j=1}^{N_s} \ \sum_{\{n\}} \ \big\lvert e^{iR(k_x \cos \theta_j + k_y \sin \theta_j)}e^{\frac{2\pi i}{N_s}j n}\big\rvert ^2
\end{equation}
where we introduced $(R, \theta_j)$ as the polar coordinates of the j-th site of the ring and decomposed ${\bf k} = (k_x, k_y)$. Using  that $\theta_j = \frac{2\pi}{N_s} j$, and defining $J_n (x,y) \doteq \sum_{j=1}^{N_s} e^{iR(k_x \cos (\frac{2\pi}{N_s} j) + k_y \sin (\frac{2\pi}{N_s} j))}e^{\frac{2\pi i}{N_s}j n}$ the previous expression can be further simplified:
\begin{equation}
 \braket{\hat n({\bf k})}\propto \frac{1}{N_s} \sum_{\{n\}} \ \big\lvert J_n(k_x,k_y) \big\rvert ^2.
\label{tof}
\end{equation}
The functions $J_n(x,y)$ provide  the discrete version of the Bessel function of order $n$. A remarkable property of such function is that $\lvert J_n(x,y) \rvert ^2$ is non-zero in the origin only when $n=0$, as shown in Fig.\ref{J_n}.

\begin{figure}
\includegraphics[width=0.7\textwidth]{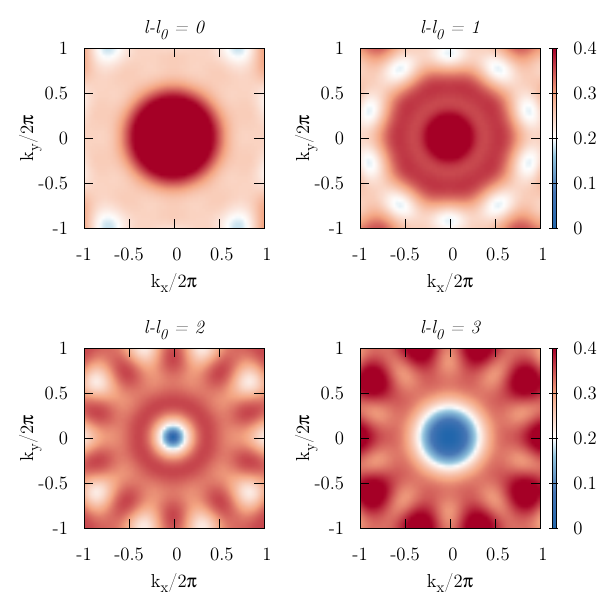}
\centering
\caption{Time of flight for $N=4$ non interacting fermions on $N_s=10$ sites. We observe the central peak to disappear when $l-l_0 \ge 2$ ie after a shift in the energy levels equal to half of the size of the Fermi sphere.}
\label{Fig_tof}
\end{figure}

From this expression it is evident that the structure of the momentum distribution strongly depends on the set of quantum numbers $\{n\}$ and therefore on the flux acting on the system. In particular, in the ground state and for $\tilde{\Omega} = 0$ such numbers are integers selected in order to minimize the current quantum number $l_0 = \sum_{\{n\}} n$, as shown in the example for $N=4$ in the right panel of Fig.\ref{J_n}. The current quantum number $l$ of the excited states is obtained by increasing each $n$ of one unit.

The momentum distribution for fermions on a lattice at $\Omega/\Omega_0 = 0$ is expected to show a peak in the center $k_x = k_y = 0$, that disappears for large values of flux \cite{PhysRevLett.95.063201}.

In order to observe the vanishing of the momentum distribution for $k_x = k_y = 0$, we have to excite all the energy levels of the Fermi sphere until $n=0$ and correspondingly $J_0(k_x,k_y)$ are excluded from the summation in Eq.(\ref{tof}). This is achieved after a shift in the energy levels equal to the half of the size of the Fermi sphere, as we show in Fig.\ref{Fig_tof} for $N=4$ particles.

\section{Continuous limit of the Hubbard ring}
We sketch the derivation of the Gaudin-Yang model as continuous limit of the Fermi-Hubbard Hamiltonian. In this section, we put $\Omega=0$ to simplify the derivation.

\begin{figure}
\includegraphics[width=0.7\textwidth]{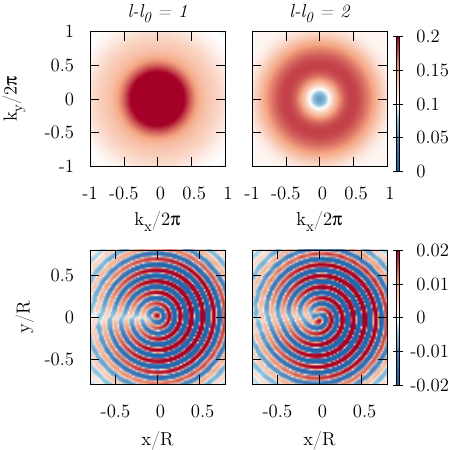}
\centering
\caption{Time of flight (top) and $\tilde{G}^{(\text{C},\text{R})}$ (bottom) for a dilute system of $N=4$ particle on $N_s=30$ sites for various values of angular momentum (left panels $l-l_0=1$, right panels $l-l_0=2$). We consider here a repulsive interaction $U/J=2$. As described in the main text, in the bottom plots we observe $N_{\uparrow}-1$ dislocations.}
\label{Fig_dilute}
\end{figure}

We define the density of fermions in the lattice as $D=N/(N_s\Delta)$,
$\Delta$ being the lattice spacing. In the continuous limit one has that $\Delta \rightarrow 0$, which implies that the filling factor $\nu=N/N_s=D \Delta$ must be accordingly small.
In the continuous limit the fermionic operators must be rescaled: $\hat c_{i,\sigma}=\sqrt \Delta \hat \Psi_\sigma (x)$,
 $\hat n_{i,\sigma}= \Delta \hat \Psi_\sigma^\dagger (x)\hat \Psi_\sigma (x) $, $x=\Delta i$.
 Then, the Fermi-Hubbard Hamiltonian reduces to the Fermi gas quantum field theory:
$\hat {\cal{H}}_{FH}=J\Delta^2 \hat {\cal H}_{FG} -2 N$,
$
\hat {\cal H}_{FG}=\int dx \left [(\partial_x \hat \Psi_\sigma^\dagger)(\partial_x \hat \Psi_\sigma)
+ c \hat \Psi_\uparrow^\dagger \hat \Psi_\downarrow^\dagger \hat \Psi_\downarrow \hat \Psi_\uparrow \right ]
$, 
with $c=U/(J\Delta)$.
The quantum fields obey the anticommutation relations $\{\hat \Psi_\sigma (x),\hat \Psi_{\sigma'}^\dagger (y)\}=\delta_{\sigma,\sigma'}\delta(x-y)$ and
$\{\hat \Psi_\sigma^\dagger (x),\hat \Psi_{\sigma'}^\dagger (y)\}=0$.
The Fermi gas field theory is the quantum field theory for the Gaudin-Yang model. Such statement can be demonstrated by writing the eigenstates of
$\hat {\cal H}_{FG}$ as $|\psi({\mathbf{\lambda}})\rangle =\int d {\mathbf z} \chi({\mathbf z}|{\mathbf{ \lambda}}) \hat \Psi_\sigma^\dagger (z_1)\dots \hat \Psi_\sigma^\dagger (z_N)|0\rangle$.
Then, it can be proven that $\chi({\mathbf z}|{\mathbf{ \lambda}})$ must be eigenfunctions of the Gaudin-Yang Hamiltonian
\begin{equation}
\hat H_{G-Y}=-\sum_{j=1}^{N_s} \frac{\partial^2}{\partial z_j^2}-2c\sum_{N_s\ge j> k \ge 1} \delta (z_j-z_k).
\end{equation}
Both the Hubbard and the Gaudin-Yang models are integrable by Bethe Ansatz\cite{essler2005one}. This should be contrasted with the bosonic case, in which the lattice reguralization spoils the integrability of the continuous theory\cite{amico2004universality}, except for the case of two bosons \cite{polo2020exact,PhysRevA.90.043606}. The mapping sketched above makes our results for  dilute lattices applicable to a one-dimensional fermionic gas with delta-interaction.

\begin{figure}
\includegraphics[width=0.7\textwidth]{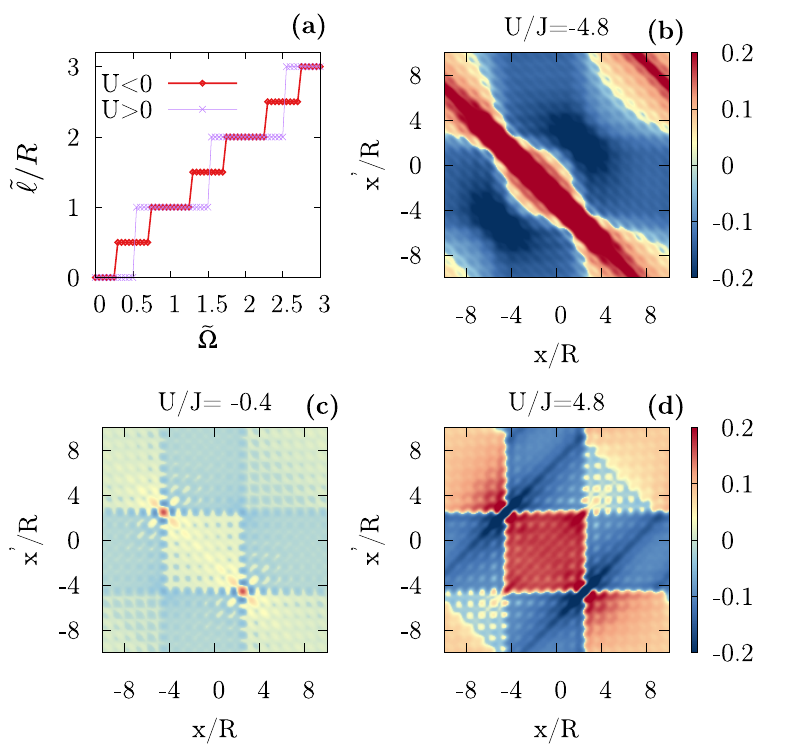}
\centering
\caption{The panel (a) shows the shift of the antidiagonal $x = -x'$ as a function of the applied flux $\tilde{\Omega}$  for $U/J=-5.4$ and $U/J=0.5$. In the panels (b), (c), (d),  we show the one dimensional correlator $G_{\uparrow,\downarrow}(x,{x'};t)$ Eq.~\ref{1d_corr} at long expansion time $t_{exp}$ and with flux $\tilde{\Omega} = 1.4$. We observe that similar structures as in the two dimensional case emerge. The peaks surfacing at $U=0.4$ are located at $k=\pm k_F$.   
The figure refers to a system of for $N=14$ and $N_s=20$.
}
\label{1d_Figure}
\end{figure}

In Fig.~\ref{Fig_dilute} we show the time of flight and the interference term $\tilde{G}^{(\text{C},\text{R})}$ at intermediate times for a dilute systems of $N=4$ particles on $N_s=30$ sites for $U=2$. We observe the characteristic hole in the center of the momentum distribution due to the action of the flux, as described in the previous section for non-interacting fermions. Furthermore, in the second line, we recognize a dislocated spiral-like interference pattern, carrying information about the angular momentum of the system and marking the fermionic nature of the particles.

\section{One-dimensional correlations}
In this section, we study the correlation function in a one-dimensional system in the reciprocal space. We shall see that we such correlations display similar features of the noise correlations in the time-of-flight momentum distribution. The correlation functions reads 
\begin{equation}
G_{\uparrow,\downarrow}^{(1d)}(x,x';t)\propto \sum_{l,j,m,n=1}^{N_s} e^{i\frac{x}{R}(l - j) + i\frac{x'}{R}(m-n)} \langle \hat c_{l,\uparrow}^\dagger \hat c_{j,\uparrow} \hat c_{m,\downarrow}^\dagger \hat c_{n,\downarrow}\rangle
\label{1d_corr}
\end{equation}
where the indices $i,j,l,m$ label the sites of the chain and $R = 2\pi L$ is the radius of the ring, $L$ being the length of the chain. Remarkably, the correlation matrix $\langle \hat  c_{l,\uparrow}^\dagger \hat c_{j,\uparrow} \hat c_{m,\downarrow}^\dagger \hat c_{n,\downarrow}\rangle$ is the same as the one used in the main text. We see in Fig.\ref{1d_Figure} that such correlator carries the same information as the two-dimensional one, with a peak along the anti-diagonal $x = -x'$ for negative $U$ revealing the formation of BCS-like pairs in the system. Because of the different geometry of the system with respect to the two-dimensional case, we observe a symmetry breaking along the diagonal $x=x'$ for non-zero flux. This is represented by a unidirectional shift of the anti-diagonal as a function of the flux that, as shown in Fig.\ref{1d_Figure}, is still occuring in quantized steps.

The connection between BCS pairing and the peaks along the antidiagonal of the noise correlator descends from the general expression of the BCS ground state. Consider indeed the state $\ket{BCS} = \prod_{k} (u_k + v_k \hat b^\dagger_{k,\uparrow} \hat b^\dagger_{-k, \downarrow})\ket{0}$, where $\hat b^\dagger_{k,\rho}$ is the creation operator for a particle with momentum $k$ and spin $\rho$ and $\ket{0}$ is the BCS vacuum. By an explicit calculation we find the noise correlator to be:
\begin{equation}
\sum_{\rho, \sigma}\braket{ \hat n_\rho(q) \hat n_\sigma(q')}= 2\abs{v_q}^2\abs{u_{q'}}^2\Bigl( \delta(q-q')\delta_{\rho,\sigma} + \delta(q+q')(1 - \delta_{\rho,\sigma}) \Bigr),
 \label{bcs}
\end{equation}
where the momentum density operator is defined as
$\hat  n_\rho(q)= \hat b^\dagger_{q,\rho}\hat b_{q,\rho}$. From this expression we see that, when $\rho \neq \sigma$, this correlation function shows a characteristic peak along $q=-q'$, with an envelope mediated by the amplitudes of the BCS state. We can also explicitly study the dependence of such correlator on the flux $\tilde{\Omega}$. To this purpose, we introduce the flux using the Peierls substitution on the creation/annihilation operators in the real space ie $\hat b_{j,\rho} \to e^{i\tilde{\Omega}j}\hat b_{j,\rho} \doteq \tilde{\hat b}_{j,\rho}$. The Fourier transform yields $\tilde{\hat b}_{q,\rho}= \frac{1}{N_s}\sum_{j=1} e^{i q j} \ \tilde{\hat b}_{j,\rho} = \frac{1}{N_s}\sum_{j=1} e^{i (q+\tilde{\Omega}) j}\ \hat b_{j,\rho} = \hat b_{q+\tilde{\Omega}, \rho}$. Therefore, the Peierls substitution shifts all the momenta of the same flux-dependent amount. This trasformation does not affect the term along the diagonal $q=q'$ in Eq.(\ref{bcs}), while the second term, proportional to the peak along $q=-q'$, get shifted by $2\tilde{\Omega}$. The stepwise dependence of the ground state on $\tilde{\Omega}$ is reflected on the noise correlator by the discrete shift of the antidiagonal we show in Fig.~(\ref{1d_Figure}a).

\section{Visibility $V(U,N)$ vs  quasi-ODLRO}
The concept of off-diagonal long range order  allows to quantify uniquely the properties of condensation or pairing in interacting quantum fluids \cite{RevModPhys.34.694}. For paired fermions, such as superconductors and atomic Fermi superfluids, the relevant correlator is the two-body density matrix:
$\rho_2(x,x',y,y')=\langle \hat \psi_\uparrow^\dagger(x) \hat \psi_\downarrow^\dagger(x')  \hat \psi_\downarrow(y) \hat \psi_\uparrow(y')\rangle$,  $\hat \psi_\sigma$ being the corresponding field operator,
which quantify phase coherence  between one pair centered at $X=(x+x')/2$ and a second one at $Y=(y+y')/2$. Existence of ODLRO implies that
$\rho_2(x,x',y,y')\simeq\lambda_0(N) \Phi_0^*(x,x')\Phi_0(y,y')$ with $\Phi_0$ natural orbital of $\rho_2$ with macroscopic eigenvalue $\lambda_0=O(N)$. Similarly, by analogy with bosonic systems \cite{colcelli2018deviations}, for quasi-ODRLO one expects that quantum fluctuations reduce the scaling to $\lambda_0(N)=O(N^\alpha)$ with $0<\alpha<1$. All the above relations are readily generalized to lattice systems, by considering the discretized versions of $\rho_2$ and $\Phi_0$ according to 
$\rho_2(j,l,m,n)=\langle \hat c_{j,\uparrow}^\dagger  \hat c_{l,\downarrow}^\dagger \hat c_{m,\downarrow} \hat c_{n,\uparrow}\rangle$ and $\Phi_0(j,l)=\Phi_0(x_j,x_l)$.
We show here how the  $G_{\uparrow\downarrow}$ correlator at long times and its visibility  allows precisely to address off-diagonal long-range order.

We start by recalling that (see Section on Time of Flight above and \cite{altman2004probing}) the long-time density correlators yield information on the correlations in momentum space of the gas before expansion:
 $ \lim_{t\rightarrow \infty} G_{\uparrow\downarrow}({\mathbf r},{\mathbf r'},t)\propto \braket{\hat n_\uparrow({\mathbf k}) \hat n_\downarrow({\mathbf k'})} $
with $\mathbf{k}= m \mathbf{r}/\hbar t$, and where the momentum correlations are defined as
\begin{equation}
 \braket{\hat n_\uparrow({\bf k}) \hat n_\downarrow({\bf k'})} =\sum_{jlmn} e^{i{ \bf k} \cdot ({\bf x}_j-{\bf x}_l)} e^{i {\bf k'} \cdot ({\bf x}_m-{\bf x}_n)} \langle \hat c_{j,\uparrow}^\dagger \hat c_{l,\uparrow}  \hat c_{m,\downarrow}^\dagger \hat c_{n,\downarrow}\rangle
\end{equation}

Hence, they are readily related to the two-body density matrix upon commutation of the order of the operators. If (quasi)ODRLO holds, along the antidiagonal $\bf{ k'}=\bf{\mathbf -k}$ one has
\begin{equation}
 \braket{\hat n_\uparrow({\bf k}) \hat n_\downarrow({\bf -k})} \simeq \lambda_0(N) |\tilde \Phi_0(\bf{k})|^2 
\label{eq:odlro}
\end{equation}
where we have defined $\tilde \Phi_0({\bf k})=\sum_{j,l} e^{i{\bf k} \cdot ({\bf r}_j-{\bf r}_l)}  \Phi_0(j,l)$ and used that $\Phi_0(j,l)$ is an even function of $j-l$. 

The connected part of the correlator $\braket{ \hat n_\uparrow({\bf k}) \hat n_\downarrow({-\bf k})}$  contains essentially the same information of ODLRO as the full one since no ODRLO is found for fermionic systems in the disconnected terms \cite{RevModPhys.34.694}.

\begin{figure}[]
    \centering
    \includegraphics{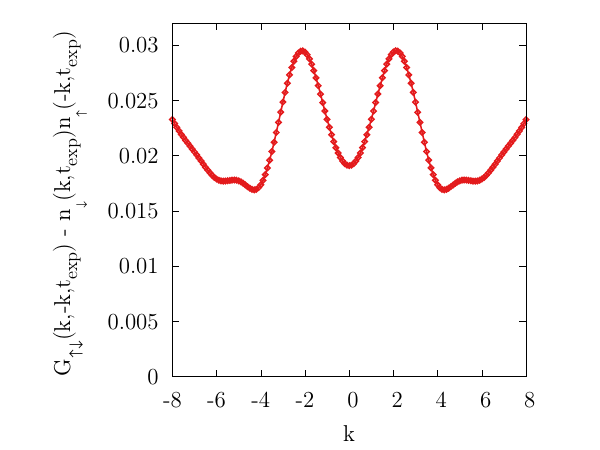}
    \caption{$G_{\uparrow\downarrow}(k,-k',t_{exp}) - n_\downarrow(k,t_{exp})n_\uparrow(-k,t_{exp}) $ for $N=14$, $N_s=20$ and $U=-4.8$. The visibility is calculated following the definition given in Eq.(7) of the main text. }
    \label{profile_vis}
\end{figure}

Combining all the above considerations, we readily conclude that the visibility of the momentum correlator along the anti-diagonal $\bf{ k'}=\bf{\mathbf -k}$,  in presence of quasi-ODLRO is predicted to behave as
\begin{equation}
V(U,N)=  \frac{\text{Max}|\tilde \Phi_0({\mathbf k})|^2 -\text{Min}|\tilde \Phi_0({\mathbf k})|^2 }{\text{Max}|\tilde \Phi_0({\mathbf k})|^2 +\text{Min}|\tilde \Phi_0({\mathbf k})|^2 }  
\end{equation}
ie, noticeably, it is independent on $N$. The visibility is then calculated using Eq.~(7) of the main text. In Fig.~\ref{profile_vis} we provide an example of profile $G_{\uparrow\downarrow}(k,-k',t_{exp}) - n_\downarrow(k,t_{exp})n_\uparrow(-k,t_{exp}) $ that we used to compute the visibility of the correlations among particles with opposite momenta.

\begin{figure}
\includegraphics[width=1\textwidth]{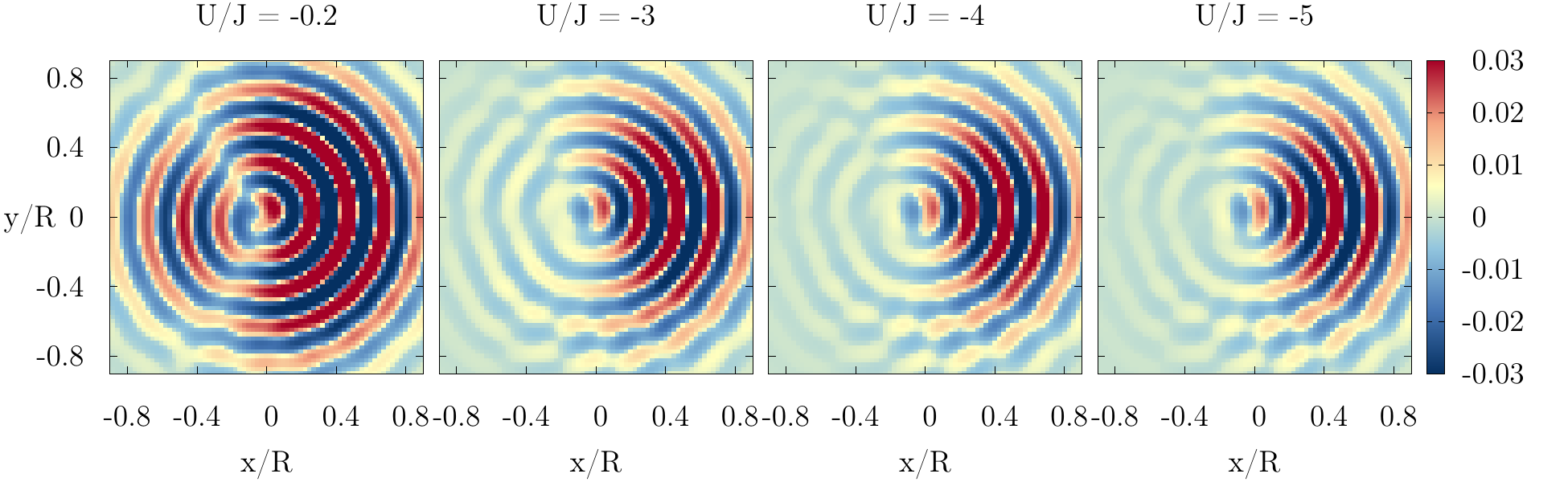}
\centering
\caption{Interference pattern $\tilde{G}^{(\text{C},\text{R})}$ for a gas of $N=6$ fermions on a lattice of $N_s = 20$ sites at different interaction strength. In each panel, the value of the flux is $\tilde{\Omega} = -1.4$. We see that increasing the interactions, the dislocations progressively gets more blurred. 
}
\label{fig_disl_vs_U}
\end{figure}

\section{Dislocations vs interactions}
In this section, we support our claim about the dependence of the dislocations in the fermionic interferograms on interaction strength. In Fig.~\ref{fig_disl_vs_U} we show that as interaction strength increases, the dislocations are less marked and precise. Since the  dislocations are related to interference among single-particle orbitals, they are and more blurred  at increasing interactions, when the system cannot be described in terms of independent particles.

\end{document}